\documentclass[sigconf,nonacm]{acmart}

\usepackage{graphicx} 
\usepackage{multirow}
\usepackage{todonotes}

\title{Listening for Expert Identified Linguistic Features: Assessment of Audio Deepfake Discernment among Undergraduate Students}

\author{Noshaba N. Bhalli, Nehal Naqvi, Chloe Evered, Christine Mallinson, Vandana P. Janeja}\thanks{Corresponding authors: mallinson@umbc.edu,vjaneja@umbc.edu; Authors would like to acknowledge NSF award \# 2210011 }
\affiliation{%
	\institution{University of Maryland, Baltimore County}
	\country{ }
}

\begin{document}

\begin{abstract}
In this paper we evaluate the impact of training undergraduate students to improve their audio deepfake discernment ability by listening for expert-defined linguistic features. Such features have been shown to improve performance of AI algorithms; here, we ascertain whether this improvement in AI algorithms also translates to improvement of listeners’ perceptual awareness and discernment ability. With humans as the weakest link in any cybersecurity solution, we propose that listener discernment is a key factor for improving trustworthiness of audio content. In this study we determine whether training that familiarizes listeners with English language variation can improve their abilities to discern audio deepfakes. We focus on undergraduate students, as this demographic group is constantly exposed to social media and the potential for deception and misinformation online. To the best of our knowledge, our work is the first study to uniquely address audio deepfake discernment through such techniques. Our research goes beyond informational training by introducing targeted linguistic cues to listeners as a deepfake discernment mechanism, via a training module. In a pre-/post- experimental design, we evaluated the impact of the training across 264 students as a representative cross section of all students at the University of Maryland, Baltimore County, and across experimental and control sections. We specifically evaluate whether there is any benefit to providing sociolinguistics-informed training to improve undergraduate students’ knowledge of, familiarity with, and discernment ability regarding audio deepfakes, and whether these effects differ by major (computing and non-computing students) and by gender. Findings show that the experimental group showed a statistically significant decrease in their unsurety when evaluating audio clips and an improvement in their ability to correctly
identify clips they were initially unsure about. While results are promising, future research will explore more robust and comprehensive
trainings for greater impact.

	\end{abstract}
    
	\maketitle
	
\section{Introduction}

Spoofed audio, specifically deepfakes, generated by or manipulated using Artificial Intelligence (AI), can lead to deception, fraud, and mis-/ disinformation. We give two real scenarios: 
First, in 2020, a bank manager in Hong Kong authorized a \$35M transfer based on a fraudulent, AI generated voice \cite{hongkong2020}. Second, in 2021, fraudsters impersonating YouTube executives arranged a phone call  with  Goldman Sachs to close a \$40M investment deal \cite{smith}. In both cases, automated fake audio detectors were not present, and listeners were left to their own discernment ability. In the  Goldman Sachs example, the team sensed that the call \textit{"sounded strange...as though it might have been digitally altered"}. Suspecting that the call was fake, their own discernment led to the fraudsters' arrest. This incident suggests that humans can not only be made aware of deepfakes, but also trained to better discern the veracity of audio they encounter.

Current research on detecting deepfakes has focused primarily on labeling content as fake and then refining algorithms, leaving human perception of deepfake audio understudied.
One recent work \cite{mai2023warning} included informational training (i.e. explaining the concept of deepfakes) to raise individuals’ awareness about deepfakes, but results were mixed and did not significantly improve listeners’ ability to actually discern deepfake audio---in either English or Chinese. Another study \cite{muller2022human} assessed listeners' capability to detect real versus fake audio using an online game; results showed that the AI algorithms performed better than or as well as the humans, who learned as they played the game but received no direct training. 

We have developed an instructional module that goes beyond  informational training, by introducing listeners to spoken English audio cues as a deepfake discernment mechanism. Two sociolinguistics experts on our interdisciplinary team selected five cues after listening to 344 fake and real audio samples; we call these \textit{Expert-Defined Linguistic Features (EDLFs)} (discussed in section 2).  Our previous work found that incorporating EDLFs improved the performance of AI algorithms in detecting fake audio \cite{khanjaniISS}, indicating the utility of incorporating domain expert knowledge for deepfake discernment. 

In this work, we thus aim to assess whether a training module using these EDLFs as foundational discernment strategies would increase listeners' awareness and perception of spoken English features in ways that improve their ability to discern fake audio.  To the best of our knowledge, our work is the first to use sociolinguistics-informed techniques to address listeners' audio deepfake discernment. Our paper shares results from a Fall 2023 training (across control and experimental groups) with 264 undergraduate students, a representative cross section of all students at UMBC, and a group frequently exposed to the potential for deception and misinformation online. 
We pose the following research questions: 
\begin{itemize}  
\item \textbf{RQ1:} Does  sociolinguistics-informed training about English audio cues improve students’ ability to discern audio deepfakes? 
\item \textbf{RQ2:} Do training effects differ for students by gender (male vs. female)? 
\item \textbf{RQ3:} Do training effects differ for students by major (computing vs. non-computing)?
\end {itemize}  

\section {Study Methodology}

\begin{figure}[h]
	\centering
	\includegraphics[trim=0cm 8cm 8cm 0cm, clip=true,scale=0.4]{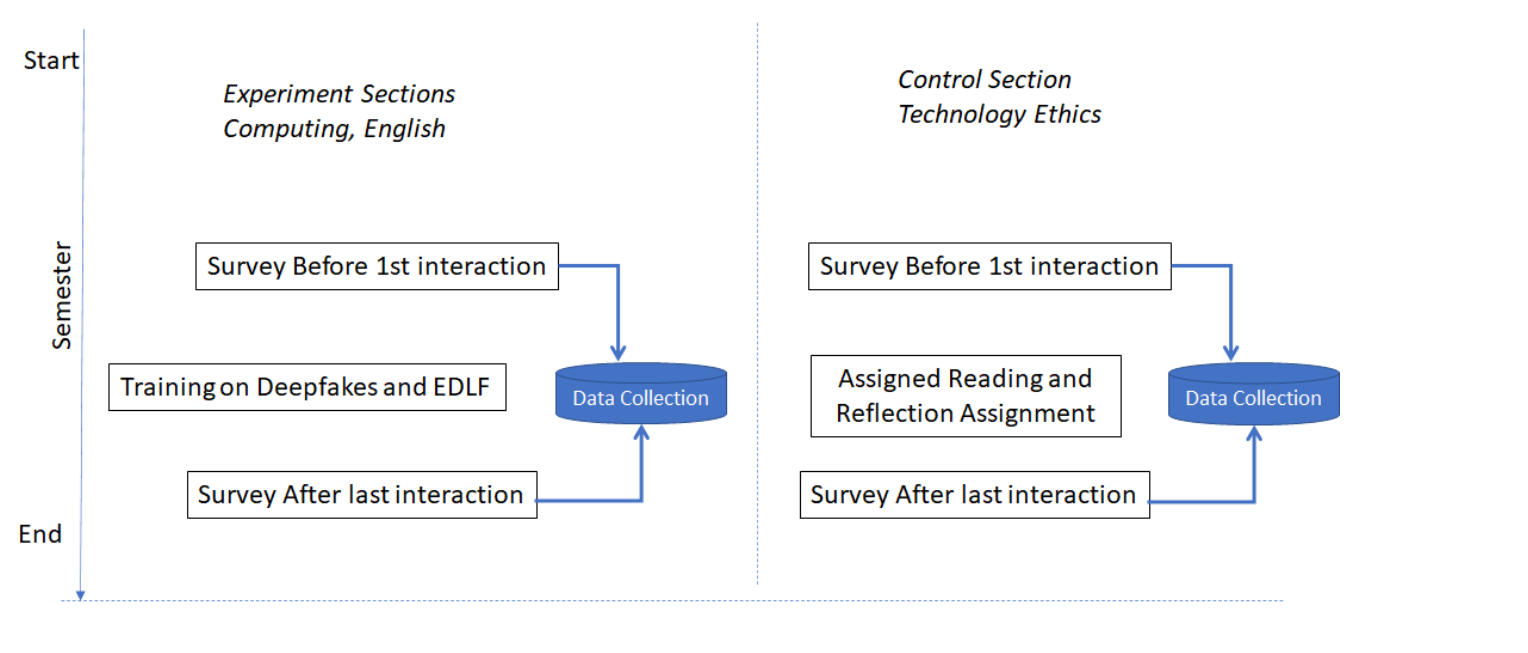} 

	\caption{Pre and post survey comparison}\vspace{-.2in}
	\label{fig1}
\end{figure}

We study the impact of training students to listen for distinguishing phonetic and phonological features of real and fake English audio – which we call Expert-Defined Linguistic Features (EDLFs), as they were selected based on the knowledge of sociolinguistics experts – as a means of discerning deepfakes. Our research introduces these targeted sociolinguistics cues to listeners as a deepfake discernment mechanism, via an instructional  module. In a pre-post experimental design in three phases, as explained below, we evaluated the impact of the training for students in experimental and control sections. 
The phases of the study design are illustrated in Figure \ref{fig1}.

\subsection{Pre-Survey and First Interaction}
In the first phase, a doctoral student leading the training explained the research study and obtained informed consent from students wishing to participate. Those students were then given an online pre-survey via Qualtrics that they completed on their own laptops using their own headphones. In this pre-survey, students were asked to listen to 20 audio clips (4 real audio, 16 fake audio) of up to 10 seconds in duration that were randomly selected from several commonly used machine learning datasets that were used in this project \cite{kinnunen_asvspoof_2017, reimao_for_2019, LJSPEECH, kumar_melgan_2019, 9746139, oord_wavenet_2016} to create the hybrid dataset. Students were asked to listen to each clip up to three times and then indicate their determination of whether the clips were real, fake, or whether they were unsure. Several open-ended questions  asked about their familiarity with audio deepfakes. 
\subsection{Training Development and Delivery}
One month later, the doctoral student returned to the classes that were part of the experimental group to deliver the training module, which taught students about the EDLFs that had been identified as useful cues in audio deepfake discernment. In comparison, classes that were part of the control group were given an  online article \cite{blue_who_2022} to read that provided an overview of deepfakes, focusing on audio.

\subsubsection{Training Module Development} 
Prior to the development of our training module, two co-authors (one faculty member, one doctoral student) with backgrounds in variationist sociolinguistics engaged in a discovery phase, reviewing a subset of 344 English audio samples from existing datasets commonly used in machine learning \cite{khanjaniISS}, in order to determine features for which the human voice audio files demonstrated perceptual variation, divergence, alteration, or absence compared to fake audio files. The sociolinguists selected five phonetic and phonological features—which we called “Expert-Defined Linguistic Features” (EDLFs)—that are frequent, easily definable, and easily discernible in spoken English: pitch, pause, word-initial and word-final release bursts of English stop consonants, audible intake or outtake of breath, and overall audio quality. These EDLFs are explained below. 

In a second phase, the sociolinguists applied binary labels to indicate where anomalous production or presence of the five EDLFs occurred in real and fake audio clips. Machine learning co-authors then incorporated these labels into supervised spoofed audio detection methods to augment AI models, finding that augmenting the audio data with expert-informed linguistic annotation increased the accuracy of spoofed audio detection significantly across the models. This finding motivated us to determine whether receiving training that familiarizes students with EDLFs could improve students’ awareness of and ability to discern audio deepfakes. 
\\
\textbf{Pitch:} Pitch was defined for this study as the perceived relative high or low tone of the speech sample. Students were advised during the training session to listen for any occurrences of pitch that may be anomalous—for example, pitch that is unusually higher or lower than expected, or unusually fluctuating or inconsistent. 
\\
\textbf{Pause: }Pause was defined for this study as a break in speech production within a speech sample. Students were advised during training to listen for any occurrences of pausing that may be anomalous – for example, lack of a pause where one would be expected, addition of a pause where one would not be expected (such as between words of a phrase), or an overly long or short pause.
\\
\textbf{Word-initial or word-final stop consonant bursts: }For this study, the sociolinguists considered the production of word-initial or word-final stop consonants, specifically the sounds /p/, /b/, /t/, /d/, /k/, and /g/ in English. Students were advised during training to listen for any occurrences of consonant release bursts that may be anomalous: for example, lack of a burst of air where one would be expected, the addition of a burst of air where one would not be expected, or an unusually exaggerated or truncated burst.
\\
\textbf{Audible intake or outtake of breath:} Students were advised during training to listen for any occurrences of intake or outtake of breath as a potential indicator of real speech. 
\\
\textbf{Audio quality: }Finally, students were advised during training to listen for each sample’s overall audio quality. This overall estimation included listening for any anomalous audio qualities-for example, sound with disturbance or distortion; sound perceived to be unusually tinny, robotic, or compressed; and the like.

\subsubsection{Training Module Phases}
After determining the sociolinguistic features to be included in the training, we designed a short, 15 to 20 minute sociolinguistically informed instructional module that aimed to improve undergraduate students' abilities to discern real from fake audio. The module consisted of four parts. The first part provided an explanation of various types of deepfakes and where students may encounter them in their lives (e.g., on TikTok). The second part introduced the research project and its methods. The third part presented each of the five EDLFs, defining them and discussing their utility as listening cues that can potentially indicate real versus fake speech. In the fourth part, examples of audio clips were played for students to practice listening for these cues. 

\subsection{Post-Survey and Final Interaction}
A month later, the doctoral student returned to each class a final time and asked students to complete the same original assessment, this time as a post-survey. At the conclusion of the semester, members of the research team visited each class to debrief with the students about the findings and their overall experiences participating in a research study; no data were collected during the debrief.

\section{Experiments and Results}

\subsection{Test Population}
In Fall 2023, we selected nine introductory undergraduate course sections based on the familiarity of these instructors with our research and their willingness to present the study opportunity to their students. Of these nine sections, two were introductory English classes and seven were introductory Computing classes. Six sections received the sociolinguistics-informed training module, while three sections served as a control group, received only a short reading passage about deepfakes, as noted above. 

A total of 264 students were enrolled across all classes. 
A total of 129 students completed the pre-survey, instructional training modules (experimental sections) or the reading about deepfakes (control sections), and post-survey. Remaining  students either opted not to participate in the study, did not complete all three parts of the study, or failed in the attention testing survey questions; in these cases, their data were not included. Results were collected across particular demographics of interest  (gender identification, native speaker status, fluency in languages other than English, computing major status) and true clip type (real or fake). Table \ref{table:table1} column \textbf{N} shows the number  of students across groups.  Here mean paired difference in accuracy is the difference in the proportion of clips that a given student answered correctly, with unsure responses also counting as incorrect.

\subsection{Findings}
To test the significance of students' changes in discernment accuracy and unsurety from pre- to post-survey, we applied a paired t-test across groups of participants.  
In cases where the distribution of differences in accuracy rates between the pre- and post-surveys did not follow a normal distribution, we moved to the next appropriate test, applying a paired Wilcoxon signed-rank test. Lastly, in cases where the distribution of differences in the accuracy rates between pre- and post-surveys was not symmetric, we used the final possible test: the paired sign test, which analyzes the signs (+ or -) of the difference in accuracy rates between the pre- and post-surveys.
Table \ref{table:table1} shows the average paired difference in accuracy for real clips, average paired difference in accuracy for fake clips, and average paired change in unsurety across each group. We present the analysis of the results below. 

\begin{table}[]
\begin{tabular}{|ll|l|l|l|l|l|}
\hline
\multicolumn{2}{|c|}{Group}                                                                                          & \multicolumn{1}{c|}{N} & \multicolumn{1}{c|}{All Clips} & \multicolumn{1}{c|}{Real} & \multicolumn{1}{c|}{Fake} & \multicolumn{1}{c|}{Unsurety} \\ \hline
\multicolumn{1}{|l|}{\multirow{2}{*}{All Students}}                                                              & C & 32                     & 0.04*                          & 0.1*                      & 0.03                      & -0.02                         \\ \cline{2-7} 
\multicolumn{1}{|l|}{}                                                                                           & E & 99                     & 0.01                           & 0.03                      & 0                         & -0.02*                        \\ \hline
\multicolumn{1}{|l|}{\multirow{2}{*}{Female}}                                                                    & C & 7                      & 0.04                           & 0.14                      & 0.01                      & 0.04                          \\ \cline{2-7} 
\multicolumn{1}{|l|}{}                                                                                           & E & 28                     & 0.01                           & 0.08                      & -0.01                     & -0.05                         \\ \hline
\multicolumn{1}{|l|}{\multirow{2}{*}{Male}}                                                                      & C & 22                     & 0.05                           & 0.08                      & 0.04                      & -0.03                         \\ \cline{2-7} 
\multicolumn{1}{|l|}{}                                                                                           & E & 69                     & 0.01                           & 0.01                      & 0                         & -0.02                         \\ \hline
\multicolumn{1}{|l|}{\multirow{2}{*}{\begin{tabular}[c]{@{}l@{}}English First \\ Language\end{tabular}}}         & C & 23                     & 0.06*                          & 0.1                       & 0.05                      & -0.02                         \\ \cline{2-7} 
\multicolumn{1}{|l|}{}                                                                                           & E & 83                     & 0.01                           & 0.02                      & 0                         & -0.03*                        \\ \hline
\multicolumn{1}{|l|}{\multirow{2}{*}{\begin{tabular}[c]{@{}l@{}}English Not First \\ Language\end{tabular}}}     & C & 7                      & -0.02                          & 0                         & -0.03                     & -0.01                         \\ \cline{2-7} 
\multicolumn{1}{|l|}{}                                                                                           & E & 12                     & 0.03                           & 0.08                      & 0.01                      & -0.03                         \\ \hline
\multicolumn{1}{|l|}{\multirow{2}{*}{\begin{tabular}[c]{@{}l@{}}Fluent in Another \\ Language\end{tabular}}}     & C & 17                     & 0.03                           & 0.1                       & 0.01                      & -0.01                         \\ \cline{2-7} 
\multicolumn{1}{|l|}{}                                                                                           & E & 41                     & 0                              & 0.01                      & -0.01                     & -0.02                         \\ \hline
\multicolumn{1}{|l|}{\multirow{2}{*}{\begin{tabular}[c]{@{}l@{}}Not Fluent in \\ Another Language\end{tabular}}} & C & 14                     & 0.05                           & 0.07                      & 0.05                      & -0.03                         \\ \cline{2-7} 
\multicolumn{1}{|l|}{}                                                                                           & E & 54                     & 0.01                           & 0.04                      & 0                         & -0.02                         \\ \hline
\multicolumn{1}{|l|}{\multirow{2}{*}{Computing Major}}                                                           & C & 21                     & 0.03                           & 0.13*                     & 0.01                      & -0.03                         \\ \cline{2-7} 
\multicolumn{1}{|l|}{}                                                                                           & E & 75                     & 0.01                           & 0.03                      & 0                         & -0.02                         \\ \hline
\multicolumn{1}{|l|}{\multirow{2}{*}{\begin{tabular}[c]{@{}l@{}}Non-Computing \\ Major\end{tabular}}}            & C & 8                      & 0.11*                          & 0.06                      & 0.13                      & 0.01                          \\ \cline{2-7} 
\multicolumn{1}{|l|}{}                                                                                           & E & 17                     & 0.01                           & 0                         & 0.01                      & -0.02                         \\ \hline
\end{tabular}
\caption{Mean Paired Difference (in Accuracy of Clip Recognition) Between Pre and Post Survey Administrations \\ \text{*} significant at $\alpha = 0.05$}
\label{table:table1} \vspace{-.3in}
\end{table}

\begin{table}[]
\begin{tabular}{|cl|l|l|l|l|}
\hline
\multicolumn{2}{|c|}{Group}                                                                                  & \multicolumn{1}{c|}{N} & \multicolumn{1}{c|}{All Clip} & \multicolumn{1}{c|}{Real Clips} & Fake Clips \\ \hline
\multicolumn{1}{|c|}{\multirow{2}{*}{All Students}}                                                      & C & 25                     & 0.05*                         & 0.10*                           & 0.04       \\ \cline{2-6} 
\multicolumn{1}{|c|}{}                                                                                   & E & 79                     & 0.02                          & 0.04                            & 0.01       \\ \hline
\multicolumn{1}{|c|}{\multirow{2}{*}{Females}}                                                           & C & 7                      & 0.04                          & 0.14                            & -0.01      \\ \cline{2-6} 
\multicolumn{1}{|c|}{}                                                                                   & E & 25                     & 0.02                          & -0.09                           & 0.01       \\ \hline
\multicolumn{1}{|c|}{\multirow{2}{*}{\begin{tabular}[c]{@{}c@{}}English First \\ Language\end{tabular}}} & C & 18                     & 0.07*                         & 0.10                            & 0.06       \\ \cline{2-6} 
\multicolumn{1}{|c|}{}                                                                                   & E & 67                     & 0.02                          & 0.03                            & 0.02       \\ \hline
\end{tabular}
\caption{Mean Paired Difference between pre and post survey in  accuracy of Clip recognition among 'Unsure Students' within demographics with significant decrease in Unsurety \\ \text{*} significant at $\alpha = 0.05$} \vspace{-.35in}

\label{table:table2}
\end{table}

\textbf{All Students:} Table \ref{table:table1} shows that the mean change in discernment accuracy for real vs fake audio is more positive in the control group than the experimental group, showing significance only in the control group's increase in accuracy when identifying real audio. The unsurety rate, i.e, students' selecting "unsure" for audio clips in the survey, decreased significantly among the students that received the training. Given these results, we further explored whether this significant decrease in unsurety is linked with an increase in accuracy of detecting fake audio. To do this, we selected students who selected "unsure" for at least one answer in the pre-survey and calculated mean paired difference in accuracy of this group. In Table \ref{table:table2}, results show that our training increased confidence for some students, but this decrease in unsurety did not always co-occur with an similar increase in deepfake discernment accuracy.

\textbf{Skepticism Among Experimental Group:} Within the experimental group, of the 353 clips that students marked as “unsure” in the pre-test, 85\% of those that referred to fake clips were correctly identified as fake in the post-test. However, only about 20\% of such real clips were correctly identified, leading us to conclude that training may have had an effect of making students more skeptical of the audio they heard, and thus more reluctant to label a clip as real.

\textbf{Control Group:} The control group, who did not take the training and only received a short reading about audio deepfakes, showed significant improvement in their accuracy when identifying clips: they showed a 4\% improvement overall across all clips, with real clips driving this trend. Students in the control group were significantly more accurate, by almost 10\%, in their ability to correctly identify short clips (<2 seconds) as real or fake. 

\textbf{Student Gender and Other Demographics:}  
In the control group, non-computing majors conformed to the group's overall trend of significant improvement for all clips, while computing majors showed significant improvement in identifying real clips only. In the experimental group, female students showed a disproportionately greater decrease in their unsurety after receiving the training. This trend among female students drives the overall trend we observe in the experimental group of decreased rates of unsurety post-training. Students who reported English as their first language also showed a beneficial decrease in their levels of expressed unsurety, but students’ majors and fluency in other languages did not significantly predict their discernment performance.

Our findings indicate the need for a more comprehensive and holistic approach to training for deepfake discernment. This would include a slightly longer training for linguistic cues and incorporating reading of related literature. Regardless of the type of training it was clear that such a training will add value to navigating the misinformation detection landscape that the students face. 

\section{Conclusion}
Detecting fake or manipulated speech continues to prove challenging. Importantly, our intervention to improve students' discernment of audio deepfakes was observed to have slightly different outcomes along certain social lines: for example, whether students spoke English as their first language predicted differences in their change in unsurety pre- and post-training. Results point to an opportunity to develop interventions and educational materials tailored to the particular needs of specific social groups, for example, those who have learned English as their second or other language. Furthermore, our finding that the reading passage educating students about audio deepfakes--which served as the control--had a positive, though limited, impact on students’ discernment accuracy indicates the need for digital media literacy education in tandem with efforts to improve listener accuracy based on audible linguistic cues. As deception by audio deepfakes continues to proliferate, efforts to improve the public’s ability to spot misleading content should take a holistic approach that incorporates insight from across disciplines.

\bibliographystyle{unsrt}
\bibliography{main}

\end{document}